%% file: main_arxiv.tex
\documentclass[11pt]{amsart}
\usepackage{geometry}                
\usepackage{graphicx}
\usepackage{amssymb}
\usepackage{epstopdf}

\usepackage{float}
\usepackage{subfig}
\usepackage{array}

\usepackage{makecell}

\usepackage{lastpage,fancyhdr}

\title{The Role of Gender in Social Network Organization}
\author{Ioanna Psylla$^1$, Piotr Sapiezynski$^{1,2}$, Enys Mones$^1$, Sune Lehmann$^{1,3}$\\
}    
\date{}                                      

\begin{document}
\maketitle
\footnotetext[1]{Technical University of Denmark, Kgs. Lyngby, Denmark}
\footnotetext[2]{Northeastern University, Boston, USA}
\footnotetext[3]{Niels Bohr Institute, University of Copenhagen, Copenhagen, Denmark}
\section*{Abstract}
\input{sections/abstract}
\section*{Introduction}
\input{sections/introduction}

\section*{Data}
\input{sections/data}

\section*{Personality}
\input{sections/results_personality}

\section*{Mobility}
\input{sections/results_mobility}
\section*{Networks and interactions}
\input{sections/results_interactions}

\input{sections/results_summary}
\section*{Gender prediction}
\input{sections/results_prediction}

\section*{Discussion}
\input{sections/discussion}

\section*{Materials and Methods}
\input{sections/methods}

\section*{Funding}
This work was supported by Villum Foundation, Young Investigator programme 2012, High Resolution Networks (SL) and University of Copenhagen, through the UCPH2016 Social Fabric grant (SL). The funders had no role in study design, data collection and analysis, decision to publish, or preparation of the manuscript.

\bibliographystyle{abbrv}
\bibliography{bibliography,References}

\end{document}

%% file: sections/abstract.tex
The digital traces we leave behind when engaging with the modern world offer an interesting lens through which we study behavioral patterns as expression of gender.
Although gender differentiation has been observed in a number of settings, the majority of studies focus on a single data stream in isolation.
Here we use a dataset of high resolution data collected using mobile phones, as well as detailed questionnaires, to study gender differences in a large cohort.

We consider mobility behavior and individual personality traits among a group of more than $800$ university students.
We also investigate interactions among them expressed via person-to-person contacts, interactions on online social networks, and telecommunication. 
Thus, we are able to study the differences between male and female behavior captured through a multitude of channels for a single cohort.
We find that while the two genders are similar in a number of aspects, there are robust deviations that include multiple facets of social interactions, suggesting the existence of inherent behavioral differences.
Finally, we quantify how aspects of an individual's characteristics and social behavior reveals their gender by posing it as a classification problem. 
We ask: How well can we distinguish between male and female study participants based on behavior alone? Which behavioral features are most predictive?

%% file: sections/introduction.tex
For many decades, gender differentiation has been studied as an interdisciplinary topic and within a variety of fields including psychology, social science, anthropology, history, and biology.
Existing studies have explored the nature of the existing gender differences, their origin, and impact on individuals' lives. 
How to interpret the observed deviation between women and men is subject to debate among scholars.
It is, however, universally accepted that behavioral differences are rooted in the different biological roles, and are reinforced by a society's values and cultural beliefs. 

Previous research has shown that gender-specific inequalities might originate from biological predispositions (e.g. hormones~\cite{wood2012gendered}, brain structure~\cite{reisberg1997cognition}), as well as the organization of the hunter-gatherer societies in which humans initially evolved~\cite{schmitt2008can}. 
This differentiation is subsequently aggravated by cultural/societal expectations~\cite{wood2012gendered, maccoby1974psychology}, which are likely to lead the two genders to develop and maintain their social ties in different ways~\cite{deaux1987putting}.
The study of social networks is essential for understanding how gender role influences the nuances observed in the structure and evolution of these social interactions. 
Although it does not provide answers regarding the origins of the gender differences in social behavior, it can help identify and understand these discrepancies to a larger extent.

Below, we first explore individual-level characteristics, specifically the psychological traits and mobility behavior within the cohort, noting that both these aspects have been found to relate strongly to social behavior~\cite{kalish2006psychological,cho2011friendship}.
Next, we focus on social traits. 
For the two gender-groups, we evaluate similarities and differences with respect to social network role. 
Our analysis of social networks is based on longitudinal data describing person-to-person interactions (physical proximity using Bluetooth scans), calls and text messages, and online friendships (based on Facebook communication activity). 
Finally, we use classification models to quantify the extent to which a person's gender can be inferred from their observed characteristics and behavior.

%% file: sections/data.tex
The basis of this paper is the \emph{Copenhagen Network Study} (CNS), a study focusing on nearly $800$ freshmen at the Technical University of Denmark~\cite{stopczynski2014measuring} who volunteed to donate data via Nexus 4 smartphones.
The bulk of data collection was behavioral data from from the smartphones, supplemented with data from online questionnaires and 3rd party APIs, such as the Facebook Graph API. 
The derived datasets include:
\begin{itemize}
\item Friendship graph and interactions (comments on wall posts) from Facebook,
\item Person-to-person proximity events, measured using Bluetooth,
\item Telecommuncation data (call and text message logs; only metadata, no content), 
\item Location records (based on GPS and WiFi), 
\item Questionnaires (responses to personality questionnaires, described in detail below).
\end{itemize}
This work is based on data collected between September 2013 and May 2014.
The number of active participants and the quality of their data varies over the duration of the observation.
To eliminate the effect of missing data on statistics, we calculate all indicators and network properties on a weekly basis and average for each individual. 
Participants with three active weeks or less during the nine-month period are excluded from the analysis.

After this filtering, this dataset consists of $166$ female and $601$ male students. 
In order to avoid the difference in population sizes affecting the standard deviations, we apply subsampling over the male and female population separately and calculate the distribution over the mean of the random sub-samples.

Below, unless otherwise specified, we use the following strategy to compare the two (female/male) classes. 
We we draw 1000 random subsamples, each equal to the half of the original class size from each class.
Then, we perform pairwise comparisons between subsamples.
We test the null hypothesis that the means of the two sampling distributions are identical (a two-tailed test).

In order to compare results across domains (personality, mobility, social interactions, etc) we measure the differences in distributions between the two genders using \textbf{effect size} $r$, defined as the ratio between the means of each distribution $x_1, x_2$ and the pooled standard deviation $\sigma_p$:
\begin{equation} \label{eq:effect_size}
r = \frac{\mu(x_1)-\mu(x_2)}{\sigma_{p}(x_1, x_2)},
\end{equation}
where $\sigma_{p}(x_1, x_2)$ is defined via
\begin{equation} \label{eq:pooled_std}
\sigma_{p}(x_1, x_2)=\sqrt{\frac{(|x_1|-1)\cdot\sigma^2(x_1)+(|x_2|-1)\cdot\sigma^2(x_2)}{|x_1|+|x_2|-2}}
\end{equation}
and $\sigma^2(x)$ is given by
\begin{equation} \label{eq:variance}
\sigma^2(x)=\frac{1}{n-1}\sum_{i=1}^{n}{(x_i-\mu(x))^2}.
\end{equation}

%% file: sections/results_personality.tex
In this section, we investigate how gender differences are expressed through personality metrics. Data from responses to personality questionnaires show that although there are considerable variations within a gender, differences between males and females exist in a number of traits and at every age. As part of the CNS study, we consider the following dimensions of personality, which are listed below along with the central gender-related results pertaining to that measure.

\begin{description}
	\item[Big Five.] The Big Five Inventory (BFI) is a widely used method for assessing human personality using five broad factors: \emph{openness}, \emph{extraversion}, \emph{neuroticism}, \emph{agreeableness}, and \emph{conscientiousness}~\cite{john1999big}. 
	To measure big five, we use the questionnaire developed in Ref.~\cite{john1999big}.
Previous work has consistently found women to be more neurotic and agreeable than men~\cite{costa2001gender, williams1999pancultural, rubinstein2005big, schmitt2007geographic, schmitt2008can, feingold1994gender}. 
There is less of a consensus with respect to gender differences in the remaining BFI attributes.
For instance, some studies report higher conscientiousness and openness among women, while others find men as more conscientious~\cite{rubinstein2005big,williams1999pancultural}.
Detailed description of each personality trait and reference to additional literature are provided in the section Personality traits of the Supplementary Information.
	\item[Self-esteem.] We use the definition that self-esteem is a feeling of self-worth~\cite{rosenberg1965society} and use Rosenberg's 10-item instrument to measure it~\cite{rosenberg1965society}.
	Feingold~\cite{feingold1994gender} found that males have slightly higher self-esteem than females, and Kling et al.~\cite{kling1999gender} showed that this effect increases considerably in late adolescence. 
However, other studies exist that show no significant difference between males and females with respect to self-esteem~\cite{maccoby1974psychology}.
	\item[Narcissism.] Narcissism has been previously found to be positively correlated with self-esteem~\cite{twenge2003isn,emmons1984factor}.
Here we assess Narcissism using the Narcisistic Admiration and Rivalry Questionnaire (NAR-Q), which integrates two distinct cognitive and behavioral aspects of narcissism: the tendency to approach social admiration through self-enhancement, and the tendency for an antagonistic self-defense (rivalry)~\cite{back2013narcissistic}.
The literature is consistent here: men tend to be more narcissistic than women, regardless the age and income~\cite{grijalva2015gender, foster2003individual, bushman1998threatened, farwell1998narcissistic}.
%
	\item[Stress.] Several studies have been conducted to measure stress levels among students in higher education, reporting that female students tend to have more stress (and more stressors) than male students, regardless of the instrument used for measurement~\cite{clark1986gender, dahlin2005stress, misra2000academic, misra2000college}.
In this study, we measure stress with the widely used Perceived Stress Scale (PSS)~\cite{cohen1983global}.
\item[Locus of control.] Locus of control reflects the extent to which a person perceives a reward or reinforcement as contingent on his own behavior (internal locus) or as dependent on chance or environmental control (extrernal locus)~\cite{rotter1966generalized}.
We measure locus of control using a simplified, 13 item scale proposed by Goolkasian\footnote{see: http://www.psych.uncc.edu/pagoolka/LC.html}.
A lower score corresponds to internal locus, whereas a higher score indicates external locus.
In general, the two genders have not been found to differ with respect to this psychological trait~\cite{rotter1966generalized, schultz2016theories, de1990sex, feingold1994gender}.
Lefcourt~\cite{lefcourt2013research} argued that those who are classified as having an internal locus of control not only perceive but also desire more personal control than individuals with an external locus and found a that females desired greater internal control than males.
However, women have been found to favor external control in items related to academic achievements~\cite{strickland1980sex}.
	\item[Satisfaction with life.] Satisfaction with life constitutes a judgment of one's life in which the criteria for judgment are up to the person~\cite{pavot1993review}.
We use the satisfaction with Life Scale (SWLS) instrument~\cite{diener1985satisfaction}, which has been widely used to assess subjective well-being within various groups of population.
The SWLS includes five generic statements, in which a subject must respond with a 1-7 scale, indicating the degree of agreement or disagreement.
Results regarding gender have been shown to be highly dependent on age.
Specifically, adolescent and elderly males have higher life satisfaction than females, while no observable difference is found among young adults~\cite{neto1993satisfaction,pavot1991further}. 
	\item[Loneliness.] We measure loneliness using the UCLA Loneliness Scale, a 20-item scale, in which a subject must indicate how often they feel an item characterizes them~\cite{russell1996ucla}.
Male college students have been found to be more lonely than female students~\cite{russell1996ucla, wiseman1995gender}.
It has also been shown that men are less willing to acknowledge feelings of loneliness, due to their more pronounced negative consequences of admitting to this feeling~\cite{borys1985gender, cramer1998sex}.
\end{description}

In summary, women tend to score higher with respect to negative emotionality (such as neuroticism and stress) than men, but it has been argued that this may be due to females more readily admitting to or perceiving such intense feelings.
Individualism also plays an important role in personality differences between the two genders.
In the present study, we analyze the gender effect on the aforementioned personality traits in an environment where females are the numerical minority, and within a highly specific group of individuals (students at a technical university).
The diverse dataset, however, allows us to combine the results from the questionnaires with the participants' behavior in a natural setting.

\begin{figure}[!ht]
\centering
\includegraphics[width=\textwidth]{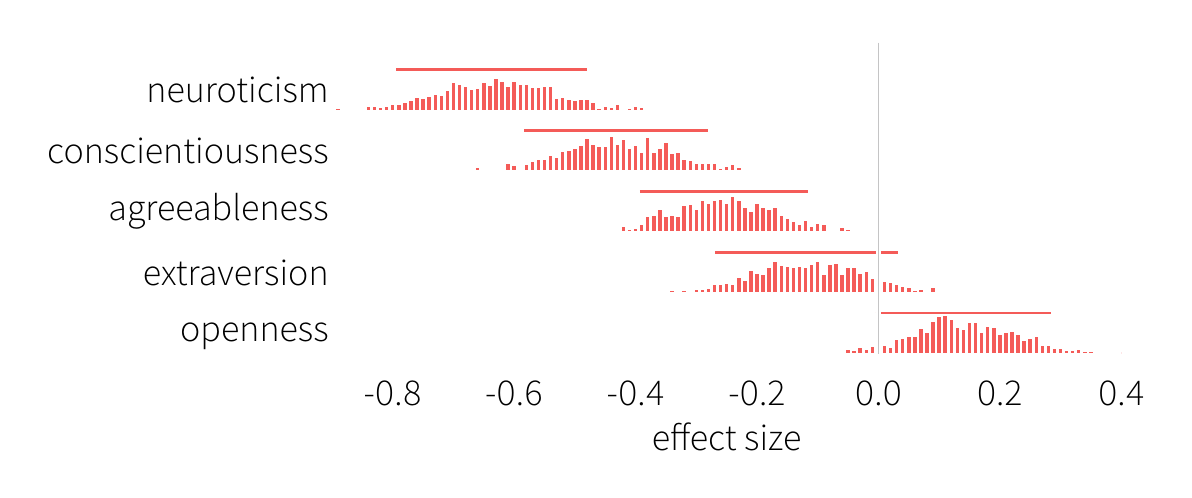}
\caption{\textbf{Signed effect sizes of the BFI measured between men and woman.}
We find female participants to be more neurotic and agreeable, in line with previous research~\cite{costa2001gender, williams1999pancultural, rubinstein2005big, schmitt2007geographic, schmitt2008can, feingold1994gender}.
Women in our study tend to be also more conscientious, and we identify no significant differences in scores for extraversion and openness. Negative values indicate that women achieve higher scores. 
Histograms show the distribution of effect sizes defined by Eq.~(\ref{eq:effect_size}), horizontal bars denote 5 and 95\% percentiles.
\label{fig:effect-sizes-bfi}}
\end{figure}

\paragraph{Results.}
We test the null hypothesis that the two samples have equal means.
We start with the Big Five Inventory, and measure effect sizes between the two genders.
Figure~\ref{fig:effect-sizes-bfi} shows the normalized difference (i.e., the effect size) observed between males and females with respect to neuroticism, conscientiousness, agreeableness, extraversion, and openness.
Each histogram represents the distribution of the difference in means normalized by the pooled standard deviation, and the mean in a subsample of females is subtracted from the mean in a subsample of males (for details on the effect size, see Eq.~(\ref{eq:effect_size})).
The horizontal bars denote $5$ and $95\%$ percentiles.
Neuroticisim exhibits the largest deviation, positioned far to the left from the zero baseline (with a mean of $d_\mathrm{neu} = -0.635$), indicating that women score significantly higher in this personality characteristic than men. 
We also find significant, albeit less pronounced, differences with respect to conscientiousness ($d_\mathrm{con} = -0.436$) and agreeableness ($d_\mathrm{agr} = -0.259$).
Finally, we do not find statistical significance in the average values of extraversion ($d_\mathrm{ext} = -0.118$) and openness ($d_\mathrm{ope} = 0.143$).
\begin{figure}[!ht]
\centering
\includegraphics[width=\textwidth]{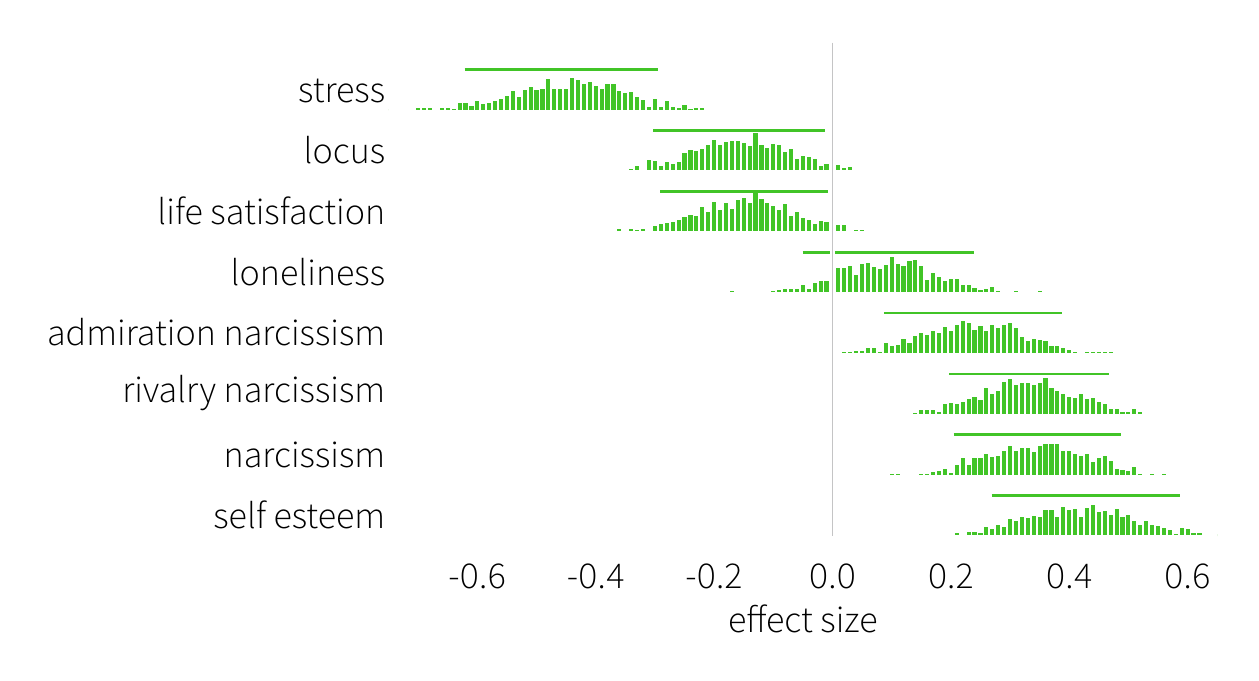}
\caption{\textbf{Signed effect sizes of personality traits measured between men and woman.}
As indicated in previous studies on college population, women tend to feel more stress than men~\cite{clark1986gender, dahlin2005stress, misra2000academic, misra2000college} and have a more external locus of control in items related to academic achievement~\cite{strickland1980sex}.
Men prove to be more narcissistic~\cite{grijalva2015gender, foster2003individual, bushman1998threatened, farwell1998narcissistic} and to have a higher feeling of self-worth~\cite{feingold1994gender,kling1999gender}.
Despite these results, and in contrast with previous literature, women in our study tend to report a higher satisfaction with life.
Negative values indicate that women achieve higher scores.
Histograms show the distribution of effect sizes defined by Eq.~(\ref{eq:effect_size}), horizontal bars denote 5 and 95\% percentiles. 
\label{fig:effect-sizes-personality}}
\end{figure}
Figure.~\ref{fig:effect-sizes-personality} depicts the results describing the remaining personality measures.
Stress is significantly higher among women ($d_\mathrm{str} = -0.451$), while it is clear that men score higher in self-esteem ($d_\mathrm{se} = 0.423$).
Overall, narcissism is higher among male students ($d_\mathrm{nar} = 0.349$), but mainly due to rivalry ($d_\mathrm{riv} = 0.334$), which is its antagonistic aspect and less because of admiration ($d_\mathrm{adm} = 0.241$), which constitutes the assertive aspect of narcissism.
We find that women score higher in I-E Rotter Scale ($d_\mathrm{loc} = -0.157$), indicating a greater average sense of external locus of control.
Women also score higher with respect to satisfaction with life ($d_\mathrm{sat} = -0.149$).
Finally, we find no statistically significant difference for loneliness ($d_\mathrm{lon} = 0.095$).

%% file: sections/results_mobility.tex
In this section we verify whether there are obsevable differences in mobility traces between the participants of the two genders. We begin by providing a brief overview on the literature discussing differences between male/female mobility patterns, then discuss findings from our cohort. 

There is a general consensus that mobility patterns are not gender neutral and womens' mobility through urban space is distinguishable from mens'.
Differences between men and women in their mobility have been ascribed to various components of the gender role, such as gender-related tasks, distinct family roles, and labor market position~\cite{law1999beyond}. 
Men and women are assumed to perform a similar number of trips, but with distance traveled and the mode of transportation differing between them.
Specifically, surveys conducted in Western countries in the '90s have demonstrated that women travel fewer kilometers than men and make more trips as pedestrians and using public transportation~\cite{polk1996swedish, rosenbloom2000trends}.
Moreover, the purpose of travel tends to differ, with women traveling most frequently for household errands and men making a majority of trips to work. 
Other studies explain the shorter commuting distances of women as a result of their weaker position in the labor market~\cite{camstra1996commuting}.
Interestingly, females have been observed to travel longer distances and explore larger areas in foraging tribes, the reason for this difference is argued to originate from the fact that women are expected to return home more frequently while gathering than men are while hunting~\cite{hanson1980gender}.

Recent studies based on mobile phone records, however, have not observed substantial differences in travel distances~\cite{kang2010analyzing}, regularity, and predictability of movements~\cite{song2010limits} between male and female commuters.
However, a study using travel diary data collected in Portland reports higher levels of activity among part-time employed women than those of part-time employed men throughout the day~\cite{kwan2000interactive}.

In conclusion, despite of recent advances in studying mobility behavior in detail based on high resolution observational data, gender-based differences are rarely observed.

\paragraph{Results. }
We follow the same procedure as in personality-related measurements: we apply subsampling to obtain equal sample sizes, calculate the effect size, and test the null hypothesis that the means of the two distributions are equal.

A common quantitative description of mobility behavior is given by the distribution of unique locations visited by an individual over some time period, e.g., using $P_u(l)$, which is the relative frequency of visiting location $l$ by individual $u$~\cite{song2010limits, sekara2015fundamental}.
Relative frequency is given by the relative time the individual spent at some location on a weekly basis.
We analyze location data obtained by periodically collecting the position estimate from the location sensor of the students' phones.
The list of unique locations that characterize an individual is extracted as a list of clusters of location measurements a DBSCAN-based algorithm developed in Ref.~\cite{cuttone2014inferring} and validated in Ref.~\cite{wind2016inferring}.

\begin{figure}[!ht]
\centering
\includegraphics[width=\textwidth]{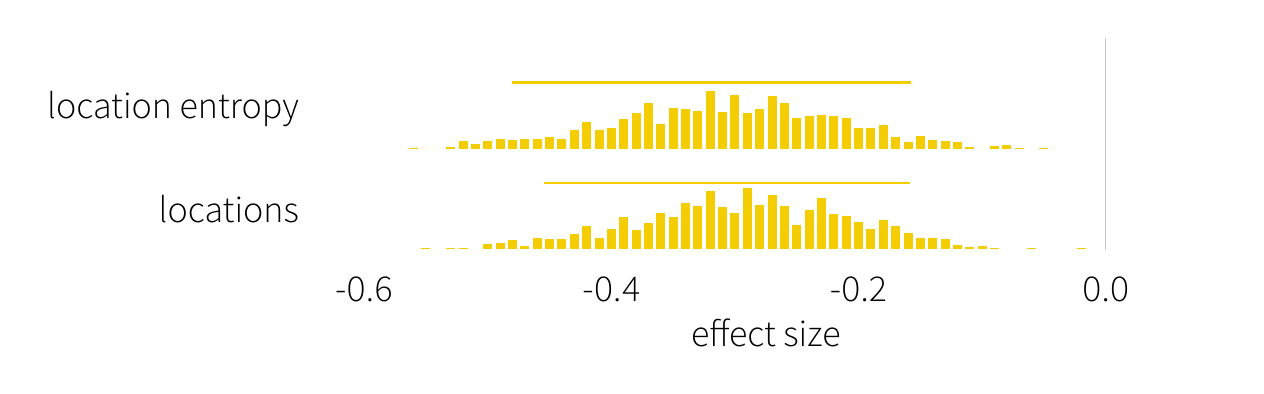}
\caption{\textbf{Effect sizes of the mobility indicators.}
We show that women are characterized by higher uncertainty in their travel patterns and visit more unique places. This deviation has not been observed in recent literature. 
Negative values indicate that women score higher than men.
Histograms show the distribution of effect sizes defined by Eq.~(\ref{eq:effect_size}), horizontal bars denote 5 and 95\% percentiles.
\label{fig:effect-sizes-mobility}}
\end{figure}

To further quantify individual mobility patterns, we measure the heterogeneity of the locations visited over time using entropy.
Entropy is a measure of uncertainty or predictability of a distribution. 
Here we use entropy to capture the heterogeneity of an individual's time spent across unique locations. 
Using $P_u(l)$, it is defined as
\begin{equation}\label{eq:mobent}
H_u = -\sum_{l \in L(u)} P_u(l) \log P_u(l),
\end{equation}
where $L(u)$ denotes the set of locations for user $u$. 
Individuals distributing their time more evenly are characterized by higher entropy.
The effect sizes measured in the location related metrics are shown in Fig.~\ref{fig:effect-sizes-mobility}.
We find that women both visit more unique locations over time, and they have more homogeneous time distribution over their visited locations than men, indicating that time commitment of females is more widely spread across places.

%% file: sections/results_interactions.tex
Now we turn our attention towards social interactions among the students. We begin by providing a brief overview on the literature discussing differences between male/female network structures. We then discuss findings based on our cohort. 

Previous work suggests that the sizes of real-world ego networks of the two genders are drawn from similar distributions~\cite{caldwell1982sex, dunbar1995social, miritello2013limited}. 
In contrast, women tend to have more friends online, as seen in multiplayer games~\cite{szell2013women} or social networking services~\cite{acar2008antecedents, sheldon2008relationship, thelwall2008social}. 
A study based on Facebook data describing around $1\,800$ U.S.~college students found that females show higher social activity and have greater betweeness centrality in their Facebook network compared to males~\cite{lewis2008tastes}.

Social networks display high gender homophily, both offline~\cite{brass1985men, onnela2014using} and online~\cite{laniado2016gender, thelwall2008social}. 
The extent of preferring same gender friends varies with age, with e.g.~girls forming smaller, more homogeneous groups than boys at young ages~\cite{mcpherson2001birds}. 
As soon as adolescents begin forming romantic ties during puberty, women start to invest more heavily in opposite-sex relationships; but they shift preferences to younger women (presumably daughters) as they age~\cite{palchykov2012sex}. 
Men, on the other hand, are shown to increase their female contacts as they get older and particularly at the end of their lifecycle~\cite{stoicaa2010age}. 
Interestingly, heterophily between genders is prominent among the strongest ties. 
For instance, calls and text conversations are both more frequent and longer among mixed-sex pairs of individuals~\cite{leskovec2008planetary,stoicaa2010age}.

Homophily has been studied as a function of transitivity (a measure of the probability of two individuals being connected provided they are both connected to the same alter)~\cite{wasserman1994social}. 
In this case, structural factors, such as network proximity, have been found to have a stronger effect on triadic closure compared to homophily \cite{kossinets2009origins}: a high number of shared contacts is a better indicator of triadic closure than sharing an attribute.
A study based on data from several U.S.~elementary schools reports that females form more triads than males and that dyads consisting of females are more likely to be in triangles~\cite{goodreau2009birds}. 
Kovanen et al.~\cite{kovanen2013temporal} studied temporal gender homophily in 3-motifs using a large dataset of mobile phone records.
They find that female-only motifs are over-represented compared to a reference model, whereas male-only motifs are under-represented.
Contradicting the aforementioned findings, however, a study based on data from the Spanish social networking site \emph{Tuenti}, found high levels of homophily in females' dyads but a higher tendency of male users to form same-sex triangles~\cite{laniado2016gender}. 

Women have not only been found to be more actively engaged in online interactions, but also to spend more time engaged in phone conversation~\cite{rakow1992gender}. 
In a review paper, Smoreda and Licoppe~\cite{smoreda2000gender} report that women tend to disclose more information to correspondents (especially about intimate topics) and are more expressive than males, which results in longer conversations, whereas men communicate mainly for instrumental purposes. 
In addition, other studies have shown that calls to a woman are longer than calls to a man regardless the gender of the caller~\cite{stoicaa2010age, smoreda2000gender}. 
Circadian rhythms in call patterns have revealed further differences between men and women, with women making longer phone calls in the evenings and during the night, and mainly when the recipient is a man (which indicates an emphasis to romantic relationships)~\cite{aledavood2015daily}. 
Likewise, young women have been reported to send a greater number of text messages, especially if the receiver belongs to the opposite gender~\cite{stoicaa2010age}. 

In summary, previous studies found clear differences in the way men and women engage with their social networks. 
However, most of the studies focus on a single channel of interactions (e.g. online communications, behavior in an organization), failing to capture a potential persistence or deviation of the characteristics across different settings. 
Here we use the CNS data to compare communication across a number of different channels. 

\begin{figure}[!ht]
\centering
\includegraphics[width=\textwidth]{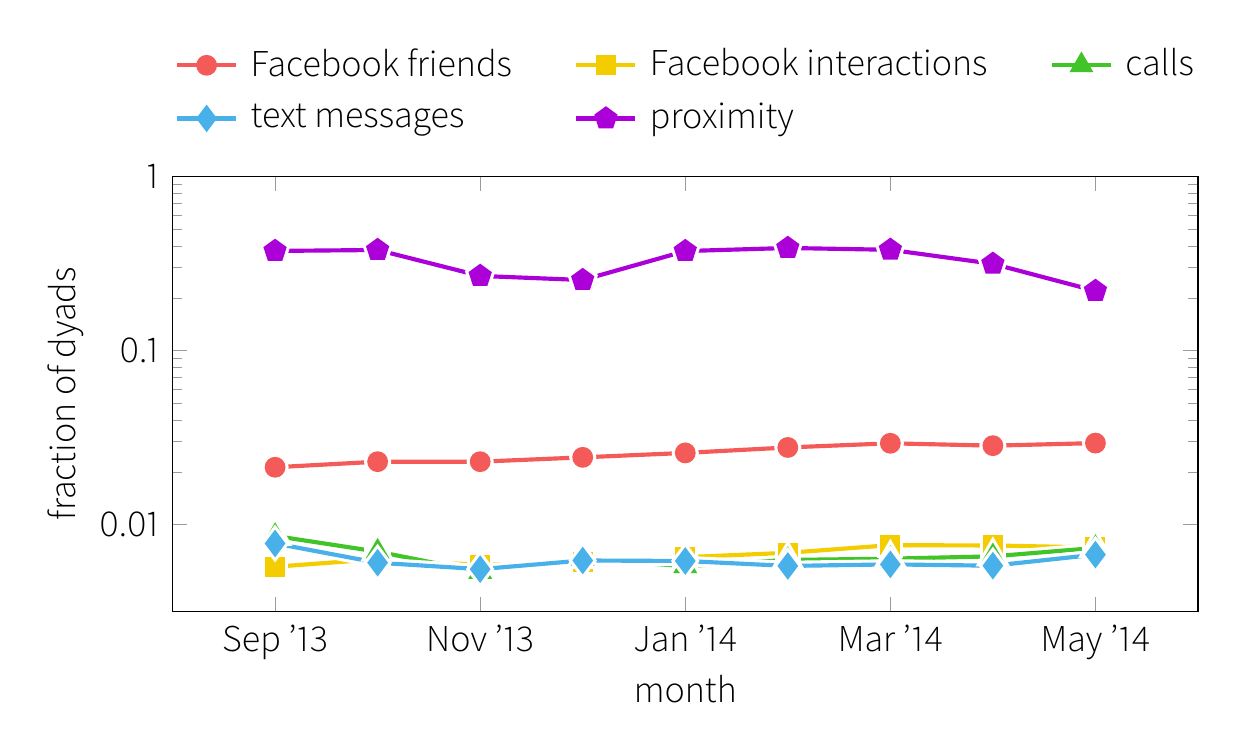}
\caption{\textbf{Fraction of active dyads in the different networks.}
Curves show the ratio between existing and potential links between participants in each network. All students attend classes at the same campus and eat at the same cafeterias, so their proximity network is very dense (with 40\% of dyads active).
Only about 2-3\% of them actually connect as friends on Facebook, and less 1\% communicate using calls, text messages, or Facebook interactions.
\label{fig:dyads}}
\end{figure}

\paragraph{Results. } 
We consider three types of communication: physical proximity (i.e., person-to-person) interactions, Facebook activity, and mobile phone communication (calls and text messages).
Previous studies have shown that each channels may describe different aspects of social ties and potentially corresponding to different levels of connection intensity~\cite{sekara2014strength, onnela2007structure, eagle2009inferring, wilson2012review, wang2013showing}.
To illustrate these differences, in Fig.~\ref{fig:dyads} we report the fraction of active links in each communication channel over time.
Note that the vertical scale is logarithmic, indicating an increased presence of proximity links (purple pentagons), a moderate level of active Facebook connections (red circles) and a comparably low level of active telecommunication links (bottom curves).
%


\begin{figure}[!ht]
\centering
\includegraphics[width=\textwidth]{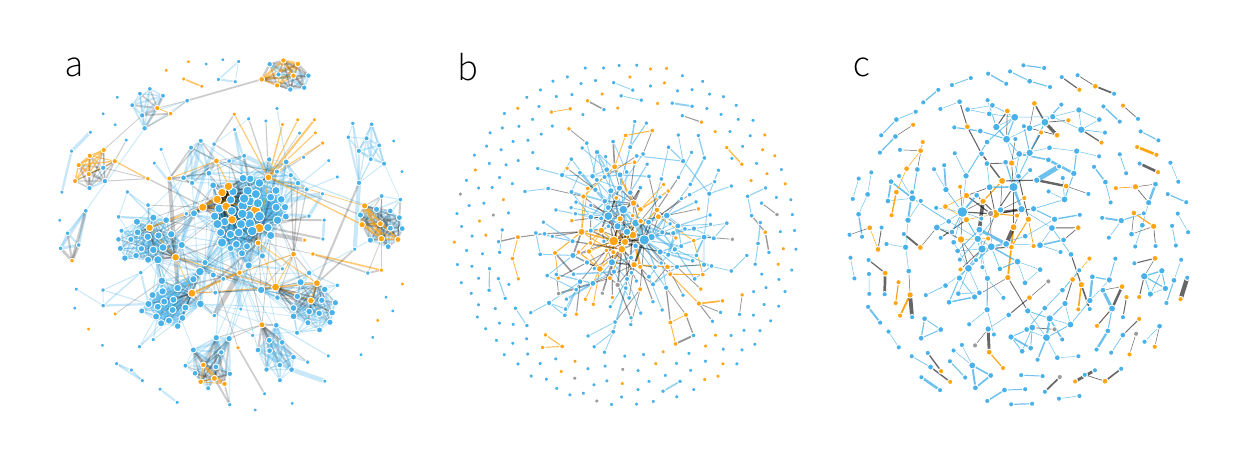}
\caption{\textbf{Snapshot of the three interaction networks: (a) person-to-person aggregated over one weekday, (b) Facebook interactions (aggregated over a month) (c) phone call networks of the students (aggregated over a month)}
Blue and orange nodes correspond to male and female students respectively, link color denotes whether the connection is between the same genders; orange for female-female, gray for mixed, and blue for male-male connections.
The size of each node is proportional to their degree and width of the links represents the frequency of interactions.
The person-to-person network shows clear separation into study groups, while this structure is no longer visible in communication networks. 
} 
\label{fig:networks}
\end{figure}
Figure~\ref{fig:networks} visualizes a single-day snapshot of the three networks in this study.
The community structure revealed in the person-to-person network is a fingerprint of the classes the students attend together.
Less structure is observed in the Facebook feed network, which shows a single large component; the call network is the most fragmented of all of the networks.
The high level of homophily in the proximity network is evident from the frequency of orange and blue lines, which represent the female-female and male-male connections respectively, as well as cliques that contain nodes of the same color. 
The call network shows the highest fraction of mixed gender connections, a possible indicator of couples.

When investigating the networks between the study participants, we apply a different approach to accounting for the imbalance male/female subjects than in case of personality and mobility.
Here, subsampling would alter the network structure and, thus, render e.g. measurements of homophily and other network metrics meaningless.
Instead, we use the following reference model: we randomly permute genders between participants with uniform probability and then perform the calculations.
Overall, we produce $2E$ network realizations, where $E$ is the number of edges in the network.

To approve or reject the null hypothesis that the network is independent of gender homophily, we calculate the  $z$-score, given by:
\begin{equation} \label{eq:zscore}
z = \frac{x - \mu(\tilde{x})}{ \sigma(\tilde{x})}.
\end{equation}
Here, $x$ is an indicator, $\mu(\tilde{x})$ and $\sigma(\tilde{x})$ are the mean and standard deviation of the indicator in the reference model.
The $z$-score is expected to be zero if the null hypothesis is true.

To test the null hypothesis of no difference between the two gender groups, we draw the permutation distribution of the differences between the two genders and measure where this distribution falls relative to the mean difference of the empirical data.
The $p$-values then are calculated by dividing the number of permuted mean differences that are larger/smaller than the one observed in the empirical data, by the number of items ($2E$) in the permutation distribution.

\begin{figure}[!ht]
\centering
\includegraphics[width=\textwidth]{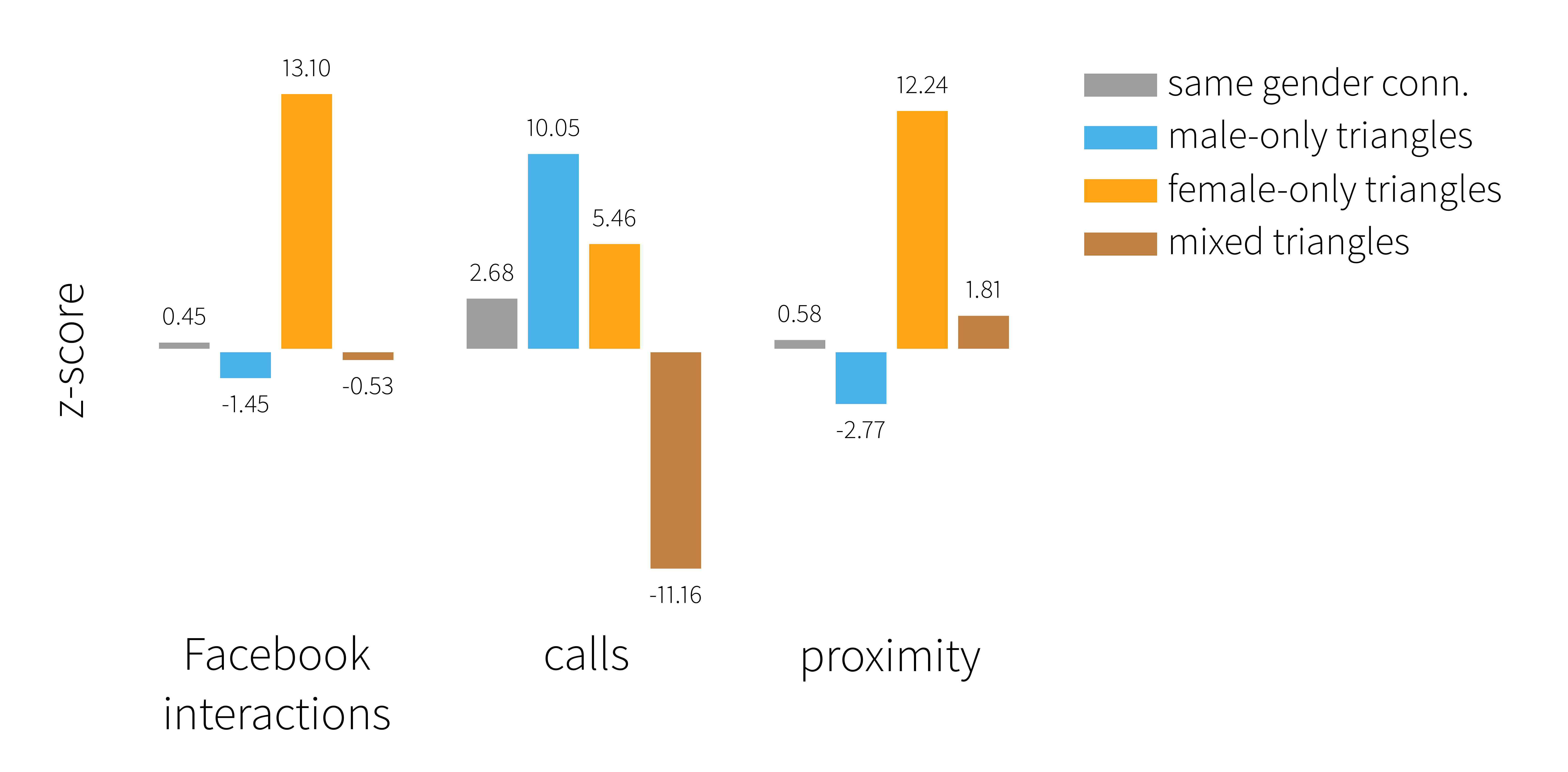}
\caption{\textbf{Homophilous connection patterns in the three interaction networks.}
Bars represent the $z$-scores of connection motifs (containing two or three individuals) obtained from the comparison with the permutation model.
Grey color corresponds to dyadic connections, color bars denote triangles: male only (blue), female only (orange) and mixed (brown). 
Women in the study are more likely to form same-gender triangles in all types of interaction networks. 
\label{fig:network_homophily}}
\end{figure}

\begin{figure}[!ht]
\centering
\includegraphics[width=\textwidth]{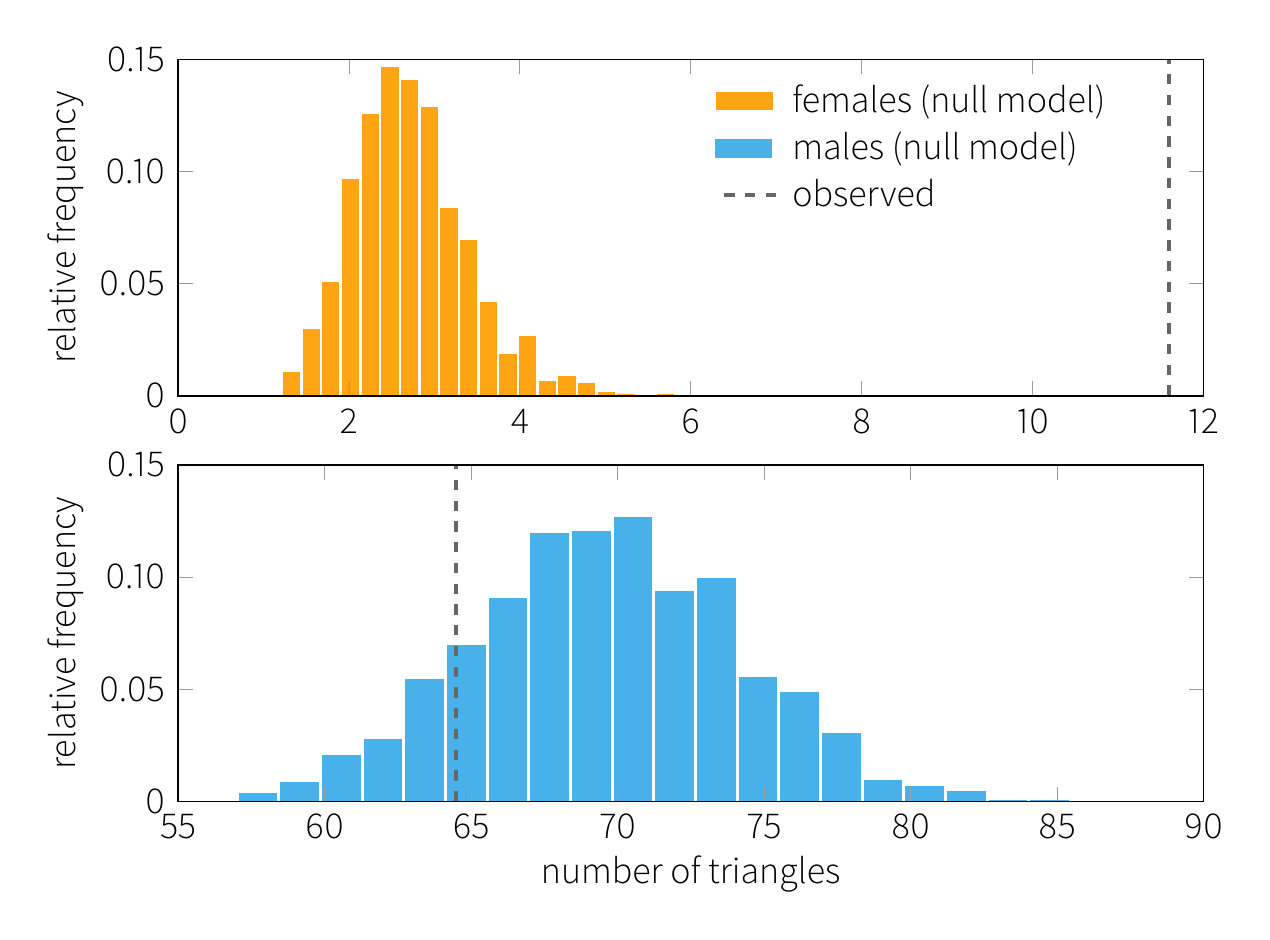}
\caption{\textbf{Distributions of same-gender triangles in the Facebook friendship network.}
Empirical observations (black dashed lines) are compared to the distribution of triangles measured in the ensemble of null models. Values on the $x$-axis correspond to the number of same-gender triangles in the network. It is noticeable that females form triangles with other females to a higher extent compared to the reference model, whereas males show the reverse pattern.
\label{fig:triangles}}
\end{figure}

We explore the influence of gender homophily on formation of friendships in the various networks among the participants. 
To do this, we first identify the fraction of same-gender friends out of all friends an individual has.
Figure~\ref{fig:network_homophily} shows the $z$-scores of various network connections obtained by comparison with the permutation model (see also Methods).
Women have remarkably more same gender friends than the ones measured with the reference model in online interactions and person-to-person interactions ($z$-score is $13.10$ and $12.24$ respectively). 
On the other hand, men also show a preference for forming homophilous ties through mobile communications, though to a less extent.
To study whether men and women tend to form closed triangles with same-gender alters, we count the various motifs in each network.
Results are shown in Fig.~\ref{fig:network_homophily} (color bars): male only (blue), female only (orange) and mixed (brown).
Furthermore, we compare the results with the respective distribution of the expected motifs found in the reference model for the Facebook network (Figure \ref{fig:triangles}). 
We find that male-only triads are insignificantly underrepresented compared to the reference model, and that there are more female-only motifs than what we would observe by chance ($z=13.101$, $p<.0001$). 
Whereas a similar pattern is observed in person-to-person interactions, same-gender motifs are overrepresented for both genders in mobile communications.
We conclude that women prefer other women for both their dyadic and triadic relationships in every form of interaction, while homophily is noticeable among males only in their trusted interactions through the phone.

\begin{figure}[!ht]
\centering
\includegraphics[width=\textwidth]{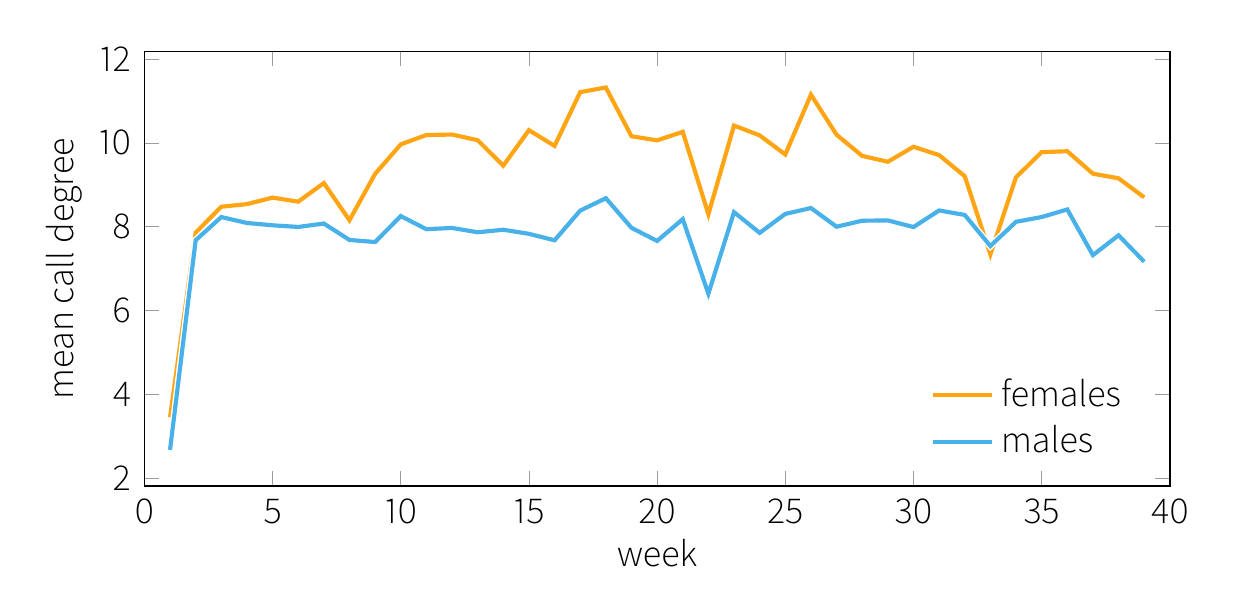}
\caption{\textbf{Degree of males and females per week in calls.}
Curves show the mean number of call contacts for males (blue) and females (orange). Women tend to have more call contact on average than men throughout the duration of the observation.
\label{fig:call-degree-trends}}
\end{figure}

We find that women tend to have a significantly higher number of contacts than men in both online and mobile networks, whereas the size of the person-to-person networks are similar.
Figure~\ref{fig:call-degree-trends} shows how degree varies over time in terms of mobile communication (calls); females have more contacts during nearly the entire period of interest.
We measure betweenness centrality (see Methods for the definition) of each individual to investigate whether one gender tends more prominently positioned in a network than the other.
We find that women consistently show higher betweenness indices, regardless of the mode of interaction. 

\begin{figure}[!ht]
\centering
\includegraphics[width=\textwidth]{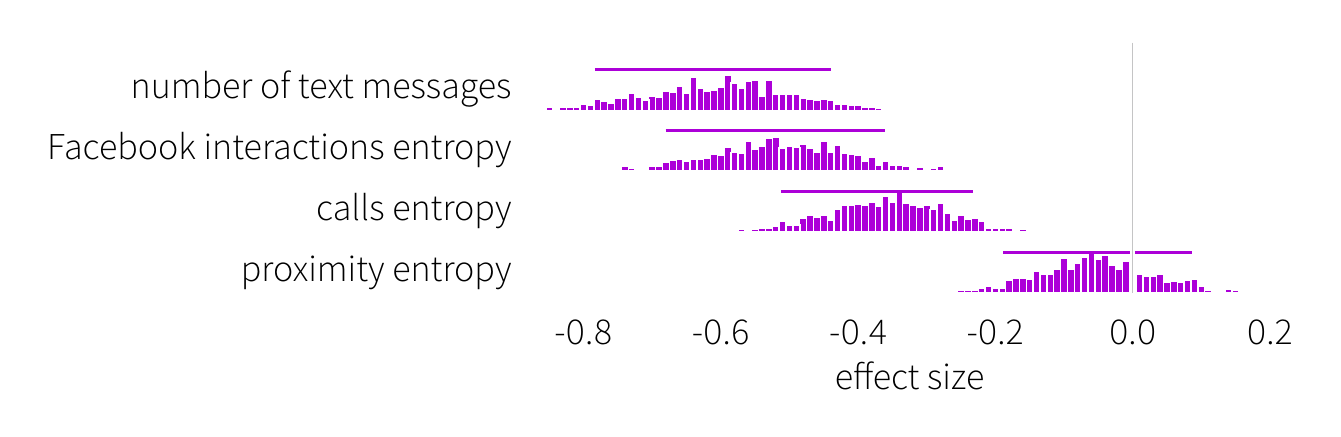}
\caption{\textbf{Effect size distribution of network metrics.}
Women tend to exchange more text messages than men and distribute their communication more evenly across their call and Facebook contacts.
We do not identify a significant difference in proximity entropy. 
Person-to-person proximity is, to a large extent, driven by class schedule and participants have less choice regarding who they interact with.
Negative values indicate that women achieve higher scores.
Histograms show the effect sizes of network related indicators (number of text messages and entropy) between males and females.
Horizontal bars denote the interval of 5 and 95\% percentiles.
\label{fig:effect-sizes-network}}
\end{figure}

Next we study the entropy of interactions.
Similarly what we did for mobility distribution in Eq.~(\ref{eq:mobent}), we calculate the entropy of the distribution of interactions over the contacts:
\begin{equation}
S_u = -\sum_{i \in N(u)} P_u(i) \log_2 P_u(i),
\label{contacts_entropy}
\end{equation}
where $P_u(i)$ is the probability that user $u$ interacts with their $i$-th contact in his ego-network $N(u)$.
The value of $S_u$ is estimated by the corresponding number of interactions relative to all interactions performed by user $u$.
Individuals who interact equal amounts with many friends will have high entropy (and therefore can be characterized by lower predictability~\cite{fano1961transmission}, whereas those who limit the vast majority of their interactions to a small set of others are expected to have low entropy (more predictable).
In Fig.~\ref{fig:effect-sizes-network}, we plot the distribution of entropy effect sizes measured between males and females for the three interaction networks. 
We observe a significant difference for Facebook and calls, with women displaying higher entropy than men, indicating that females distribute their interactions with friends considerably more homogeneously. 
In addition, females exchange remarkably more text messages than males ($p<.001$).
With respect to time spent on social interactions, we find that in our study, women are described by significantly longer conversation times during phone calls than men, regardless of the initiator of the call ($p<.0001$).
The longest calls (on average) are observed on ties where a male initiates contact to a female (with an average duration of 117 seconds), with second longest average call-durations observed between females (an average duration of 114.56 seconds). 
The shortest average duration (71.52 seconds) is measured between pairs of males.

%% file: sections/results_prediction.tex
Based on the findings presented above, we consider the classification problem of predicting gender based individual and social characteristics. In the literature, there have been attemts to predict gender based on Call Detail Record (CDR) data using semi-unsupervised techniques and deep-learning algorithms~\cite{frias2010gender, felbo2015using}.
De Montjoye et al.~\cite{de2013predicting} found that gender is a strong predictor of neuroticism, a trait that is seen in the literature to be consistently higher among women. 
In this study, we combine the questionnaire data, mobility patterns, as well as social interaction habits of each participant, to build a dataset that offers adequate complexity for achieving a good performance in the gender-inferrence problem. 
Additionally, the machine learning process also provides insight into the question: What are the most predictive behavioral indicators of gender.

\paragraph{Results.}
We use the behavioral measures calculated above as features to train four different models: \emph{logistic regressor}, \emph{AdaBoost}, \emph{support vector classifier} (SVC), \emph{random forest}, and \emph{gradient boosting classifier} implemented in the scikit-learn Python package~\cite{scikitlearn}.
Each models is evaluated using 10-fold stratified cross-validation.
Each of the models underwent a hyperparameter fine tuning procedure described in detail in the Supplementary Information.

Men constitute 78\% of the study participants. 
This poses a significant imbalance in the data and therefore, we measure performance using the area under receiver operating characteristic curve (ROC-AUC) which is robust againts imbalance, as well as using the $F_1$ score that is sensitive to the imbalance.
The value of ROC-AUC can be interpreted as the probability that the classifier is able to identify the female in of a male/female pair.
The $F_1$ score is the harmonic mean of precision (what fraction of people identified as women are actually women) and recall (what fraction of women are identified as women) at a selected threshold.

\begin{figure}[!ht]
\centering
\includegraphics[width=\textwidth]{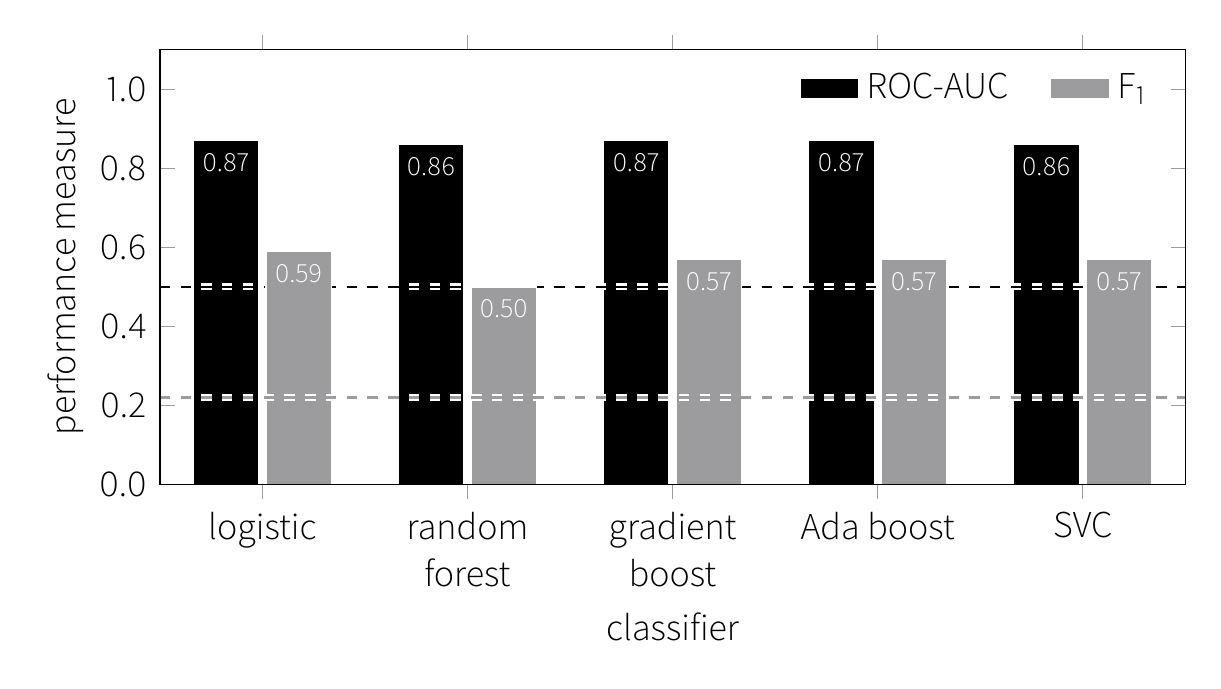}
\caption{\textbf{Classifier performance in gender prediction.}
Reported are the ROC-AUC (black) and F$_\mathrm{1}$ (grey) scores for the five classifiers considered in the gender prediction.
Dashed lines mark the baseline obtained from the random classifer that takes the imbalance into account.
High values of area under receiver operating characteristic (ROC-AUC) indicate a strong separation between men and women in the feature space: when presented with a random male and a random female, the classifier will identify the female correclty 87\% of times.
\label{fig:model_performance}}
\end{figure}

Results are summarized in Fig.~\ref{fig:model_performance} for each classifier along with the corresponding values of the random classifier based on the imbalance present in the data.
All classifiers after the hyperparameter fine tuning procedure perform similarly well, with ROC-AUC values of $0.86$ and $F_1$ scores of $0.5$ and higher (compared to a random classifier with ROC-AUC of $0.5$ and $F_1$ of $0.22$).

\begin{figure}[!ht]
\centering
\includegraphics[width=\textwidth]{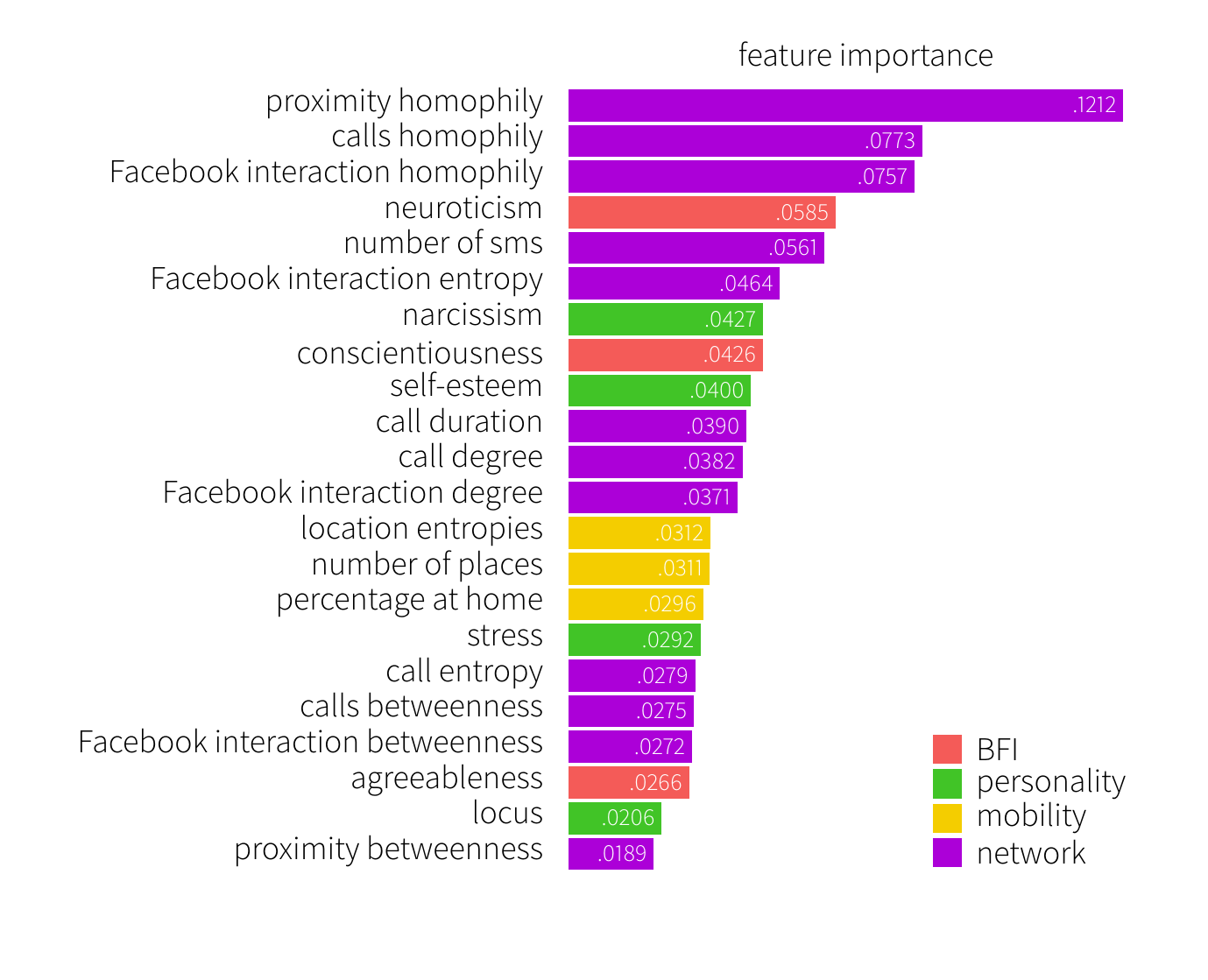}
\caption{\textbf{Feature importance obtained by Random Forest.}
Color shows the corresponding feature types: BFI (red), personality traits (green), mobility (yellow) and network (purple).
The most important features are those who help separate men from women most:
women, on average, reveal higher degree of homophily, neuroticism, and conscienciousness, while men tend to score higher in tests on narcissim, and self-esteem.
}
\label{fig:feature-importance}
\end{figure}

Next, we investigate the question of which behavioral features are most informative regarding gender.
To do this, we use the feature importance obtained by fitting a random forest model to the data (Fig.~\ref{fig:feature-importance}).
We find that that a tendency towards gender homophily in the social networks is the most important behavioral feature; this is true for all three types of interactions that we consider. 
Some aspects of personality are also important. 
Within the big five traits, neuroticism and conscientiousness are most predictive, while narcissim and self-esteem are most powerful among the remaining personality tests.
High on the list, we also find various communicatation related network characteristics.
With respect to feature types, network indicators are the most important ones, occupying five of the top six indicators.

%% file: sections/discussion.tex
\subsection*{Conclusion}
In this work we have studied gender differences within a population of freshmen. 
We have been able to identify a number of gender differences in personality traits (measured via questionnaires), mobility patterns, as well as social network behavior based on person-to-person, telecommunication, and online social networks.

\emph{Personality.} Starting from gender differences with respect to personality, our findings are in accordance with observations in the psychology literature on gender differences.
Discrepancies (or differences that are not significant) correspond to personality traits that, according to previous research, display ambiguous behavior over genders.

\emph{Mobility.} With respect to mobility behavior, our results are not consistent with findings in the literature. 
Previous work has found a restricted travel space for women~\cite{law1999beyond}, but we find that women travel more than men on average. 

\emph{Networks.} Humans use multiple channels when we communicate: real-world conversations unfold when we meet person-to-person, we call each other and send text messages, we engage with Facebook posts and write comments, send email, and use other instant messaging platforms.
Based on the communication channels we have access to here, we find that each of these channels plays a slightly different role with respect to gender similarities and differences. 
First, within all networks, we observe differences with respect to gender homophily,
specifically an over-representation of female-only dyadic and triadic connections.
This overrepresentation is mostly emphasized in weak links, which is consistent with the literature~\cite{kovanen2013temporal, thelwall2008social, reeder2003effect, szell2013women}.
Males show a lower level of homophily.
For stronger social ties (that is, contacts require more effort to maintain), both genders show similar level of homophily.
Second, we find that women tend to simply communicate more.
On average, females maintain more contacts than males in the population, they exchange significantly more text messages, and talk longer on the phone.
This is expressed both via a larger number of contacts as well as a higher entropy of neighbors.
Furthermore, in agreement with a previous study on Facebook~\cite{lewis2008tastes}, we find that, in general, women have more central positions in the network, as expressed through a higher average betweenness centrality.


\emph{Predicting gender.} Finally, we use the features discussed above to predict gender based on personality and behavioral patters.
The prediction task is based on combined personality, mobility and network features for each individual study participant, allowing us reveal the relative importance of each feature in predicting gender.
We find that personal characteristics and social behavior can be used to identify the gender of an individual with high performance ($AUC = 0.87$).
We find that network features are highly revealing, followed by personality test scores.

\subsection*{Limitations}
Our results point out significant differences in various aspets of social behavior between males and females, based on a population of nearly $800$ freshmen at a large European university.
However, in order to have a clear understanding of the results, it is important to note the limitations of the dataset as well as the methods applied, which we address in the following paragraphs.

\subsubsection*{Population sizes}
In our dataset, male population is around four times larger than the female population. 
This skewed female/male ratio presents may present certain biases simply because, in addition to gender differences, women may behave like a minority in some cases (e.g. with respect to homophily in the social networks). 
The female/male ratio also presents methodological challenges.
We describe our approach to mitigate these issues, both at the individual level as well as for the network analysis in the Methods section.

\subsubsection*{Demographics}
The cohort of the CNS experiment consists of Danish and international freshmen at a technical university.
Population impose strong constraints on demographics, with respect to age and social embeddedness.
Furthermore, we do not have detailed demographic information regarding the contacts the participants made outside the experiment.
Although demographic information is necessary to understand the results, and individuals located in Denmark indeed display different personal behavior (for example, a low overall level of neuroticism), our results regarding the comparison of males and females are in agreement with existing literature on personality traits.

\subsubsection*{Non-binary gender identification}
In this work, as in Copenhagen Networks Study in general, the participants reported their gender through a questionnaire that only offered two options: female and male.
This limiting distinction might have contributed to additional noise in the measurement of differences as well as to lowering the performance of the models in the gender-inferrence task.

%% file: sections/methods.tex


\subsection*{Network metrics}
We construct three types of networks representing the various interactions among the participants: physical proximity, Facebook, and call networks.
We then aggregate them over time windows of one week.
Only consenting participants of the CNS are included in the networks, since we do not possess complete information (e.g, gender and social activity) about the other contacts of the students.
However, extending the ego networks of the students with individuals outside the experiment (for instance, in the call networks), we can extract additional descriptors about the participating students, such as their total number of contacts or distribution of mobile phone conversation times.

In the present study we show statistics over different network metrics that are related to the local structure of the graphs and the position and role of participants in the global network.
In the following, we provide a detailed description of the applied network metrics.
\begin{description}
\item[Degree.] The number of contacts $k_i$ an individual has in their respective social network.
We calculate the degree in two different settings. First, by considering all contacts of a student, we can infer the total degree (without referring to the gender of the contacts).
Second, by limiting the interactions to the participants, we calculate the degree describing same-gender contacts.

\item[Betweenness centrality.] This measure quantifies the importance of an individual with respect to information flow on the network, when the shortest paths are taken into account.
It is defined as:
\begin{equation}
C_\mathrm{B}(i) = \sum_{j\ne k\ne i} \frac{n^\ell_{jk}(i)}{n^\ell_{jk}}.
\end{equation}
Here, $n^\ell_{jk}$ denotes the number of shortest paths between individuals $j$ and $k$ among which $n^\ell_{jk}(i)$ number of paths go through individual $i$.
Therefore, betweenness effectively measures the fraction of shortest paths that pass an individual, which is a precursor of their relevance in case of any flow on the network (rumor propagation, spread of an infectious disease, etc). 
\end{description}
\subsection*{Imputation of missing values}
Due to the method of data collection some fraction of students has missing data in various channels.
Overall, 21.5\% of the participants exhibit missing data in at least one channel.
To address the problem, we first remove participants with missing features in more than two of the five feature categories (personality, location, call, Facebook, and person-to-person interactions).
We then apply a KNN based imputation to the remaining data~\cite{troyanskaya2001missing}, described as follows.
For each user we find their $k$-nearest neighbors (with $k=7$) by calculating the average difference of non-missing features with other users.
We only use features that are present in the potential neighbor's feature set, that is, if $L_{uv}=F_u\cap F_v$ denotes the set of overlapping features of users $u$ and $v$, the distance between the users is given as:
\begin{equation}
d_{uv} = \frac{1}{|L_{uv}|}\sum_{i=1}^n |x^{(u)}_i-x^{(v)}_i|
\end{equation}
where $x^{(u)}_i$ and $x^{(v)}_i$ are the values of the $i$-th feature for users $u$ and $v$ respectively, and $|L_{uv}|$ is the size of the overlap set.
Once the $k$-nearest neighbors of all users are determined, for each student we impute their missing values by the average of the corresponding non-missing feature values of their neighbors.
If there is a single neighbor, their value is assigned.

\subsection*{Fine tuning of the machine learning models}
We fine-tuned each of the models used in the gender prediction task.
Through a grid search with cross validation we found the set of hyper parameters for which each model achieved the highest harmonic mean between $F_1$ score and ROC-AUC on previously unseen data.
Table~\ref{tab:grid_search} lists the parameter values in the grid. Optimal values are bold.
\begin{table}[h!]
\begin{tabular}{l|p{9cm}}
Model & Parameters \\ \hline
LogisticRegression & C:\{1, .2, .5, \textbf{1}, 2, 5, 10\} \\ \hline
RandomForestClassifier & \makecell[l]{max\_features:\{\textbf{all}, sqrt, log2\}, \\n\_estimators:\{\textbf{1000}\}} \\ \hline
GradientBoostingClassifier & \makecell[l]{max\_features:\{all, sqrt, \textbf{log2}\}, \\learning\_rate:\{0.001, 0.002, 0.005, .01, \textbf{.02}, .05\}, \\n\_estimators:\{\textbf{1000}\}} \\ \hline
AdaBoostClassifier & \makecell[l]{learning\_rate:\{0.01, \textbf{0.02}, 0.05, 0.1, 0.2, 0.5, 1.0\}, \\n\_estimators:\{\textbf{1000}\}} \\ \hline
SVC & \makecell[l]{C:\{.01, .05, 0.1, .5, 1, 5, \textbf{10}\},\\ kernel:\{\textbf{linear}, poly, rbf, sigmoid\}} \\ \hline

\end{tabular}
\caption{Each model used in the gender prediction task was fine-tuned. Optimal values for each parameter are marked in bold.}
\label{tab:grid_search}
\end{table}

\subsection*{Personality traits}
We consider eleven personality traits in the main paper, with five traits forming the Big Five Inventory.
Table~\ref{tab:personality_traits} lists all the personality traits along with their definition and references to the corresponding literature.

\begin{table}[h]
\begin{tabular}{l|p{9cm}}
Personality trait & Description \\ \hline
Openness* &  Describes original, independent, creative and daring behavior \cite{schultz2016theories,goldberg1981language,digman1990personality,john1999big}. \\
Conscientiousness* & Desribes how careful, reliable, hardworking and organized is one \cite{schultz2016theories,goldberg1981language,digman1990personality,john1999big}. \\
Extraversion* & Sociable, talkative, fun-loving and affectionate \cite{schultz2016theories,goldberg1981language,digman1990personality,john1999big}. \\
Agreeableness* & Good-natured, softhearted, trusting and courteous \cite{schultz2016theories,goldberg1981language,digman1990personality,john1999big}. \\
Neuroticism* & Describes worried, insecure, nervous and highly strung state \cite{schultz2016theories,goldberg1981language,digman1990personality,john1999big}. \\
Self-esteem & A 10-item instrument to define the ``feeling of self-worth''. Example items are ``I feel that I have a number of good qualities'' and ``I certainly feel useless at times'' \cite{rosenberg1965society}. \\
Narcissism & A 18-item instrument (Narcisistic Admiration and Rivalry Questionnaire) describing a ``grandiose view of the self, a strong sense of entitlement and superiority, as well as tendencies to show dominant, charming, bragging, and aggressive behaviors'' \cite{back2013narcissistic}. \\
Perceived stress & A 14-item instrument (Perceived Stress Scale) that measures ``the degree to which situations in one's life are appraised as stressful'' \cite{cohen1983global}. \\
Locus of control & A 29-item instrument (I-E Rotter Scale) to measure ``the extent to which a person perceives a reward or reinforcement as contingent on his own behavior (internal locus) or as dependent on chance or environmental control (extrernal locus)'' \cite{rotter1966generalized,lefcourt2013research} \\
Satisfaction with life & A 5-item instrument (Satisfaction with life scale) that measures ``life satisfaction as a cognitive-judgmental process''. An example item is ``In most ways my life is close to my ideal'' \cite{diener1985satisfaction,pavot1993review}. \\
Loneliness & A 25-item instrument (UCLA Loneliness Scale) describing the ``perceived loneliness'' \cite{russel1978developing,russell1996ucla}. \\ \hline
\end{tabular}
\caption{Definition of the personality traits considered in this paper for the male and female freshmen students.
Traits marked with an asterisk (*) are part of the Big Five Inventory.
References provide further reading on the various personality traits.}
\label{tab:personality_traits}
\end{table}